\pdfoutput=1
\documentclass[11pt]{article}
\usepackage[margin=1in]{geometry}
\usepackage{amsmath,amssymb,graphicx}
\usepackage{longtable,booktabs,array,calc}
\usepackage{float}
\usepackage{wrapfig}
\usepackage{needspace}
\usepackage[font=small,labelfont=bf]{caption}
\usepackage{textcomp}
\usepackage[hidelinks]{hyperref}

\providecommand{\real}[1]{#1}
\makeatletter\renewcommand\@biblabel[1]{#1.}\makeatother
\begin{document}
\begin{center}
{\LARGE\bfseries A high-speed anamorphic pupil-conjugate slit spectrograph for rapid, self-luminous sources}\\[1.1em]
{\large Matthew Vayner and Kyle D. Gilroy}\\[0.25em]
Vision Research, 100 Dey Road, Wayne, New Jersey 07470, USA\\[0.25em]
{\small matthew.vayner@ametek.com \quad kyle.gilroy@ametek.com}
\end{center}
\vspace{0.4em}
\begin{center}\begin{minipage}{0.92\textwidth}\small
\textbf{Abstract.} We present a slit imaging spectrograph whose wavelength registration is a property of the instrument rather than of the pointing of the fore-optic or the position of the source. An internal telecentric 4f relay places the system stop within the instrument while a cylindrical lens places the slit aperture at a plane where transverse position encodes source angle only (an \(A = 0\) conjugate in the ABCD matrix formulation). As a result, the lateral motion of the source or the pointing of the fore-optic does not cause the spectral image illuminating the sensor to wander, thus eliminating any non-physical registration of chemical species not present in the event and the need for on-the-fly recalibration. Additionally, the use of a high-speed CMOS camera eliminates temporally averaged spectral lines and aliasing noise. We characterize wavelength calibration (He discharge lamp, quadratic fit, RMS residual 0.12 nm, verified with Ne discharge lamp, same calibration, RMS residual 0.13 nm), spectral resolution (R $\approx$ 639, FWHM 0.92 nm at the optimal slit width setting, R $\approx$ 783, FWHM 0.75 nm at the instrument profile floor), and pointing invariance (He 587.6 nm line wander 0.29 nm RMS, 1.03 nm peak-to-peak, a $\sim$100$\times$ reduction relative to the same instrument with the aperture stop at the fore-optic pupil). We perform combustion spectroscopy on ignited nitrocellulose at 10 kHz, 99 \textmu{}s exposure time, identifying Na, K, Ca, CaOH, and Rb chemical species using the unchanged wavelength calibration transferred from the helium discharge lamp.
\end{minipage}\end{center}
\vspace{0.8em}
\section{Introduction}
Time-resolved emission spectroscopy of spatially evolving sources such as exploding wires, pulsed plasmas, and combustion/detonation fronts poses a problem\textsuperscript{1--3} for the typical slit-based spectrograph: the source moves. As the source translates, expands/contracts, or rotates, the spectral image illuminating the sensor will wander accordingly, causing unknown and apparent wavelength shifts as well as temporal averaging and aliasing noise from the motion. The latter is resolved with a high-speed camera and short exposure times, but the former features events of interest that are too fast to recalibrate on-the-fly and often you can only blow something up once.

Currently, the standard solution is fiber-coupled spectrographs which decouple source spatial and angular information from spectral output through fiber modal scrambling.\textsuperscript{4,5} However, fiber-based solutions also present several issues: the fiber bundle can only sample a discrete set of spatial points (as determined by the number of fibers), they carry per-fiber wavelength and throughput calibrations, they introduce modal noise to the spectral output, and they heavily degrade and attenuate deep-UV light.\textsuperscript{6}

An alternative is to set the spectral registration to a plane conjugate to the pupil rather than to the source (image conjugate). Definitionally, at this location, transverse position encodes ray angle rather than source position so any information about source location doesn't propagate downstream to the spectral dispersion. This solution is not new and has been arrived at independently in several communities -- in fiber double scramblers, the second fiber's input face acts as the pupil conjugate;\textsuperscript{7} in densified pupil spectroscopy, the detector itself is the pupil conjugate;\textsuperscript{8,9} LSDpol is a degenerate case where the slit \emph{is} the stop (the stop is trivially an image of itself) with field-of-view set by a field stop upstream;\textsuperscript{10} the Atmospheric Infrared Sounder uses the entrance slits as the pupil conjugate;\textsuperscript{11} pupil-slicing reformatters, as the name implies, place the pupil conjugate at the slicer;\textsuperscript{12} rod-lens slit homogenizers don't place anything at the pupil conjugate but the virtual conjugacy plane is one focal length away from the optics that the detector itself is conjugated to;\textsuperscript{13} in back-focal plane spectroscopy for nano-photonics, the slit aperture is the pupil conjugate although interestingly, angular information is of primary importance for these applications rather than a nuisance.\textsuperscript{14,15}

What has not been done, to our knowledge, is adapting this principle in a compact, free-space instrument using stock catalogue optics for the purpose of pointing invariance such that light coming from fast, self-luminous, and transient one-shot sources maintains fixed spectral positions on the sensor. Here, the slit aperture is placed at the pupil conjugate formed by an internal aperture stop such that no spatial information propagates downstream through to spectral output.

Additionally, we use anamorphic optics (a cylindrical doublet) to retain coarse spatial localization in the axis perpendicular to the wavelength (dispersion) axis. The vertical extent of each band is set by the iris diameter and therefore trades directly against throughput. As above, such an idea is not new: specifically, a patented slit-homogenizer design\textsuperscript{13} uses a rod lens with power only on the dispersion axis and is placed behind a physical slit, where the physical slit sits at the image conjugate. Across the dispersion axis, the rod lens forms a pupil conjugate one focal length downstream which the spectrometer images onto the detector (virtual slit); in the perpendicular axis it adds no optical power, so in this direction it completely preserves spatial resolution. We apply the same principle of decoupling space and spectral information using the anamorphic optic, however we place the cylindrical doublet \emph{before} the slit aperture, which is key in forming the pupil conjugate at the slit plane -- the slit-homogenizer places the slit at the image plane, such that object space movement would translate the light off the slit and would simply be vignetted.

We present the following: (i) a description of the spectrograph made with stock catalog optics and ABCD matrix analysis identifying the condition under which pointing invariance holds; (ii) characterization of the pointing invariance versus aperture stop location, demonstrating how the instrument is agnostic to choice of fore-optic; (iii) wavelength calibration and spectral resolution characterization; (iv) a demonstration on a spatially extended source with an uncontrolled combustion front at 10 kHz with species identification from an unchanged calibration.
\section{Principle of Operation}
\begin{figure}[!htb]
\centering
\includegraphics[width=0.8\textwidth]{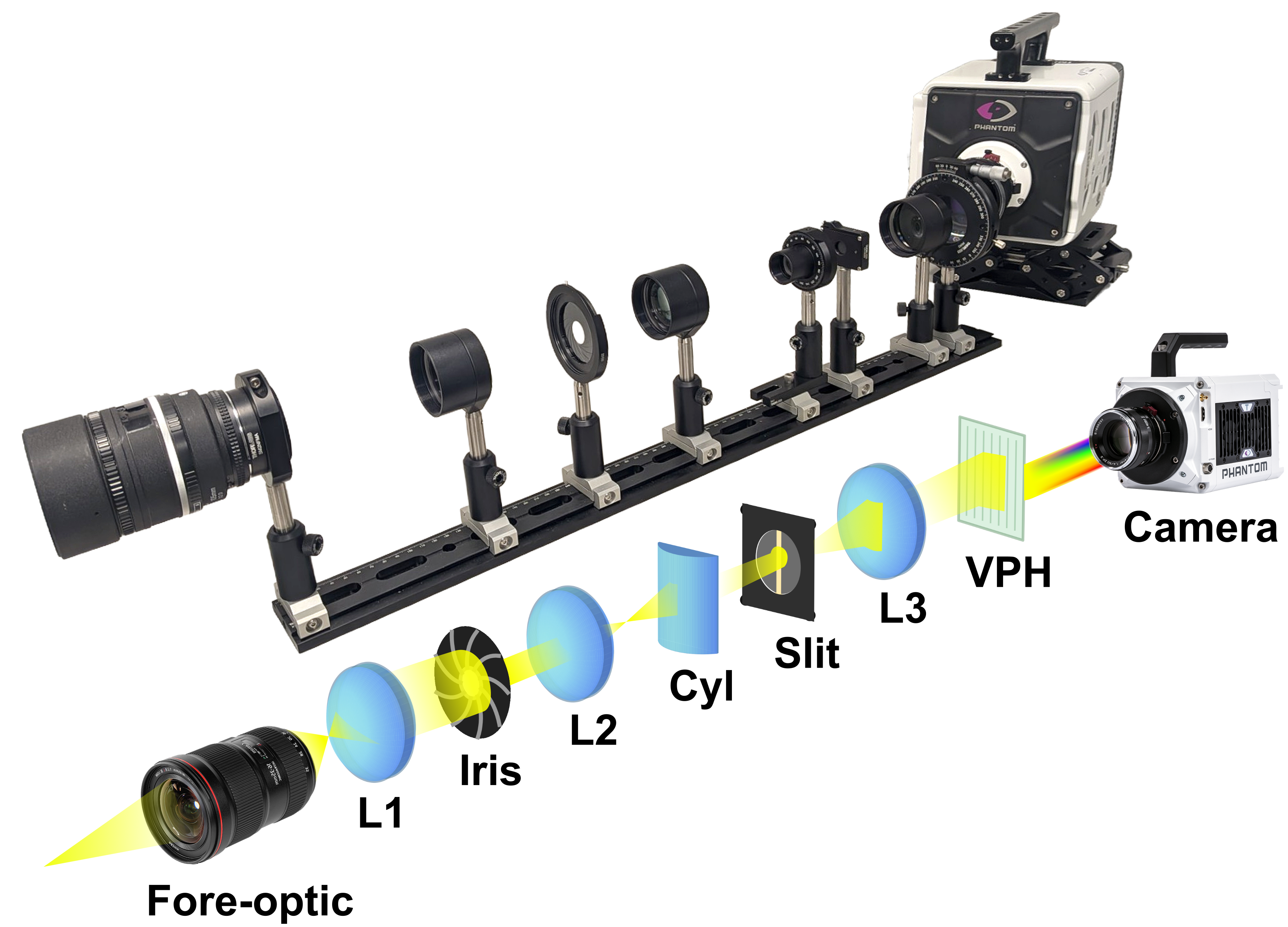}
\caption{Top: Photograph of the setup. Bottom: Diagram of the optical train.}
\end{figure}
Figure 1 shows the optical train: a photographic fore-optic, f=100mm OD=2'\,' achromatic doublet (L1), iris aperture, f=100mm OD=2'\,' achromatic doublet (L2), f=50mm, OD=1'\,' cylindrical doublet (Cyl) with power in lateral (wavelength) direction only, slit aperture, f=100mm OD=2'\,' achromatic doublet (L3), visible volume phase holographic (VPH) transmission grating, and Phantom camera with f/1.4 85mm photographic lens focused at infinity. The first three elements L1 (placed such that the front focal plane of L1 coincides with the flange focal distance of the F-mount fore-optic), Iris, and L2 form a 4f telecentric relay -- the iris is placed at the Fourier plane of the relay such that the pupil plane of the front-end (before the slit aperture) remains within the instrument and isn't shifted forward to the aperture of the fore-optic. The cylindrical doublet is arranged in a 2f (Fourier transform) configuration with its front focal plane at the image plane formed after L2 and the slit aperture placed at its back focal plane, thus placing the slit aperture at the angle-space conjugate of the source, i.e. propagating the ray vector \(\left( x_{0},u_{0} \right)\) (where \(x_{0}\) is the source position above the optical axis and \(u_{0}\) is the source angle with respect to the optical axis) from the fore-optic image through the relay and cylindrical lens along the lateral axis yields a slit-plane coordinate \(x_{s} = - f_{c}u_{0}\) where \(f_{c}\) is the focal length of the cylindrical doublet -- a quantity explicitly independent of \(x_{0}\). Additionally, \(u_{0}\) has no implicit dependence on \(x_{0}\) if the iris in the 4f relay is the aperture stop of the system. If the iris isn't the aperture stop of the system, pointing invariance is no longer guaranteed since generally \(u_{0}\) transmits the slope window \(u_{0} \in - \left( \frac{A}{B} \right)x_{0} \pm R/|B|\) where \(R\) is the half-width of the aperture stop, and \(A\), \(B\) are elements of the ray transfer (ABCD) matrix\textsuperscript{16} where any arbitrary ray arrives at the slit plane with height with respect to the optical axis \(Ax_{0} + Bu_{0}\). An aperture placed where \(A = 0\ \)therefore transmits a slope window whose center carries no dependence on source position. A complete step-by-step ABCD matrix derivation is given in the Supplementary Information. Thus, constructing the front-end of the instrument is a two-step procedure: first, place the iris in the 4f configuration and stop it down enough such that it, rather than the fore-optic aperture, is the system stop -- this removes implicit dependence on \(x_{0}\) from \(x_{s}\). Second, place the cylindrical doublet in a 2f configuration such that its front focal plane corresponds with L2's back focal plane (the image plane) and place the slit aperture at the back focal plane of the cylindrical doublet -- this removes explicit dependence of \(x_{0}\) from \(x_{s}\).

Figure 2 demonstrates the importance of the aperture stop location and the potential introduction of spectral wander: if the aperture stop shifts from the iris plane to the pupil of the fore-optic, the spectral image is no longer locked and the result is significant wander. For the three scenarios, the source (LED flashlight) was placed at the farthest off-axis left position (red) still within the field of view (FOV) of the instrument and swept across the field to the farthest off-axis right position (blue) still within the FOV (62 cm with working distance 1.5 m). In the case of the fore-optic's pupil plane acting as the aperture stop, there is significant spectral wander (RMS 65.27 col, 1.18 mm at the sensor when tracking the LED pump peak) and inconsistent light intensity at all wavelengths. Meanwhile, when the iris aperture acts as the aperture stop, even with the slit wide open, there is nearly zero wander (RMS 1.37 col, 24.66 \textmu{}m) with very consistent light intensity, demonstrating the purpose of the 4f telecentric relay. The last scenario with the slit closed shows optimal operating parameters (RMS 0.63 col, 11.34 \textmu{}m).

\begin{figure}[!htb]
\centering
\includegraphics[width=\textwidth,height=0.88\textheight,keepaspectratio]{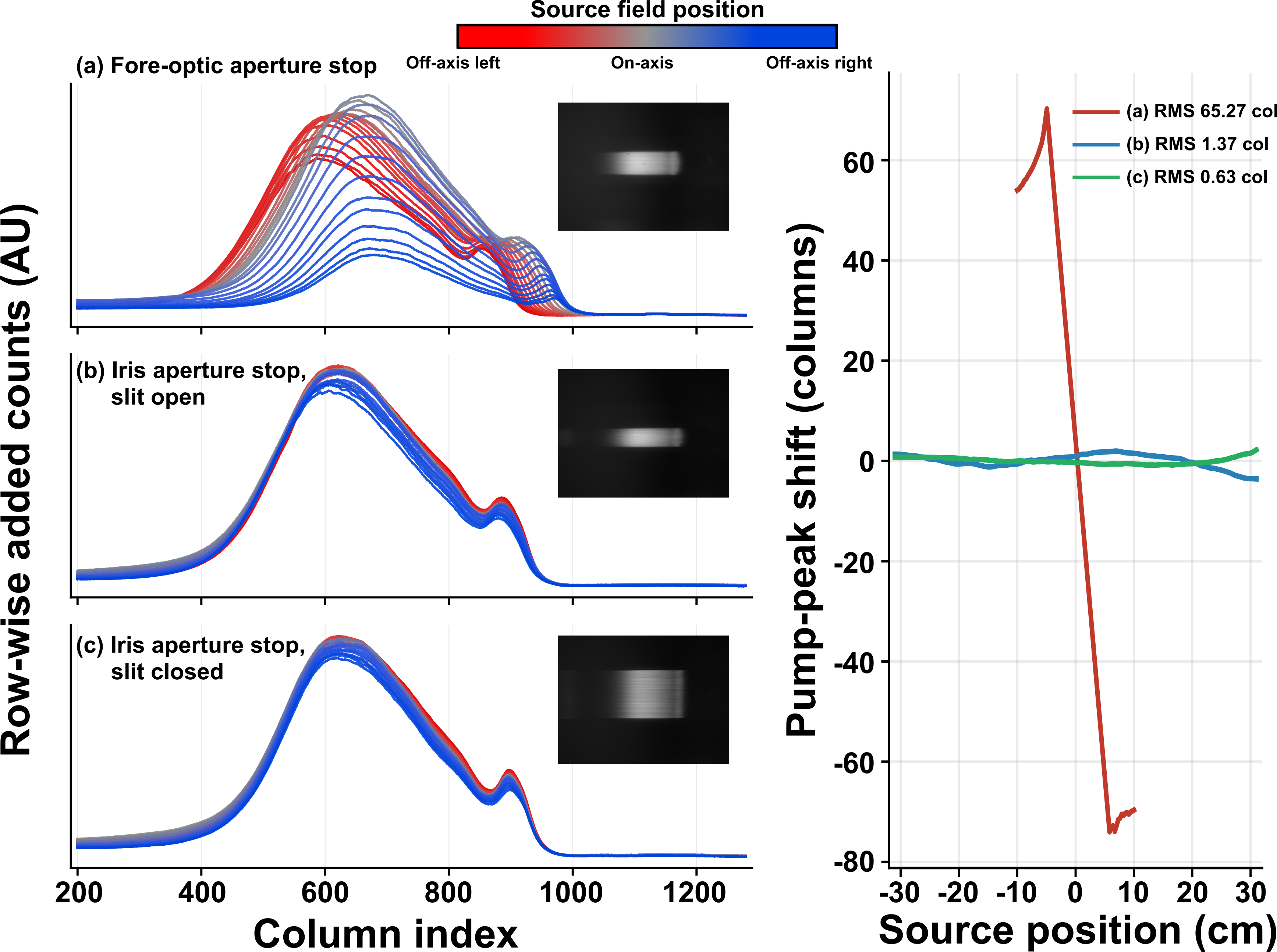}
\caption{Spectral wander and aperture stop location. (a) Fore-optic aperture acting as the system stop causes significant wander throughout the usable FOV and across all wavelengths. (b) Even with the slit aperture wide open, the iris aperture acting as the system stop maintains excellent wavelength locking ability across the usable FOV, with a slight chromatic shift near the peak of the LED output around column 600. (c) Optimal setup for the spectrograph with the iris as the system stop and slit aperture closed. Wavelength is locked to a column index with sub-col RMS and intensity ratios are preserved throughout the sweep. Right: Tracking the pump peak around column index 900 throughout the sweep. Note that (a) has significantly less FOV -- since the slit plane is no longer the pupil conjugate, light will fall off the slit opening and simply be clipped well before the case for (b) and (c).}
\end{figure}

Given that during optimal operation the fore-optic's aperture should be wide open while the user stops down the iris just enough to make it the aperture stop, the instrument remains completely agnostic to choice of fore-optic as the aperture stop will always remain at the iris plane.

Lastly, the back-end (after the slit aperture) of the spectrograph remains standard: L3 is used as a collimator placed one focal length away from the slit aperture and the VPH grating provides the wavelength-dependent dispersion necessary to measure spectral output. The camera has a photographic lens mounted on and focused at infinity as the light coming off the grating is still collimated from L3. Similar to above, it is interesting to note that since the light coming off of L3 is also a collimated bundle, the axial position of the camera does not affect the wavelength locking with the photographic lens focused to infinity, removing a degree of freedom that is the most sensitive to placement and, clearly, the most crucial to the wavelength locking mechanism.
\begin{longtable}[]{@{}
  >{\raggedright\arraybackslash}p{(\columnwidth - 2\tabcolsep) * \real{0.3267}}
  >{\raggedright\arraybackslash}p{(\columnwidth - 2\tabcolsep) * \real{0.6733}}@{}}
\caption{Instrument list.}\tabularnewline
\toprule\noalign{}
\begin{minipage}[b]{\linewidth}\raggedright
\textbf{Parameter}
\end{minipage} & \begin{minipage}[b]{\linewidth}\raggedright
\textbf{Value}
\end{minipage} \\
\midrule\noalign{}
\endfirsthead
\toprule\noalign{}
\begin{minipage}[b]{\linewidth}\raggedright
\textbf{Parameter}
\end{minipage} & \begin{minipage}[b]{\linewidth}\raggedright
\textbf{Value}
\end{minipage} \\
\midrule\noalign{}
\endhead
\bottomrule\noalign{}
\endlastfoot
Fore-optic & F-mount photographic lens (instrument is agnostic to this choice) \\
Relay L1, L2 & f = 100mm, Ø50.8mm achromatic doublets, 4f telecentric \\
Aperture stop & Ø50.8mm max iris at relay Fourier plane \\
Anamorphic element & f = 50mm, Ø25.4mm cylindrical doublet, power only in dispersion axis \\
Collimator L3 & f = 100mm, Ø50.8mm achromatic doublet \\
Grating & VPH transmission, 600 lines/mm, first order \\
Camera lens & 85mm f/1.4 focused at infinity \\
Camera & Phantom VEO 1310 \\
Spectral range & 400-820 nm \\
Spectral resolution & 0.92 nm FWHM (R $\approx$ 639) at knee; 0.75 FWHM (R $\approx$ 783) at instrument profile floor \\
Wavelength accuracy & 0.12 nm RMS (He, on-axis); 0.13 nm RMS (Ne, on-axis transferred solution) \\
Pointing invariance & 0.29 nm RMS, 1.03 nm peak-to-peak across usable FOV \\
Field of view (FOV) & 62 cm at 1.5 m working distance for pointing invariance profiling; 33.6 cm at 3 m working distance for combustion demonstration (configuration-dependent, see Discussion) \\
Demonstration acquisition & 10 kHz frame rate, 99 \textmu{}s exposure \\
\end{longtable}
\makeatletter\@nobreakfalse\makeatother
\section{Wavelength Characterization}
\begin{figure}[!htb]
\centering
\includegraphics[width=0.9\textwidth]{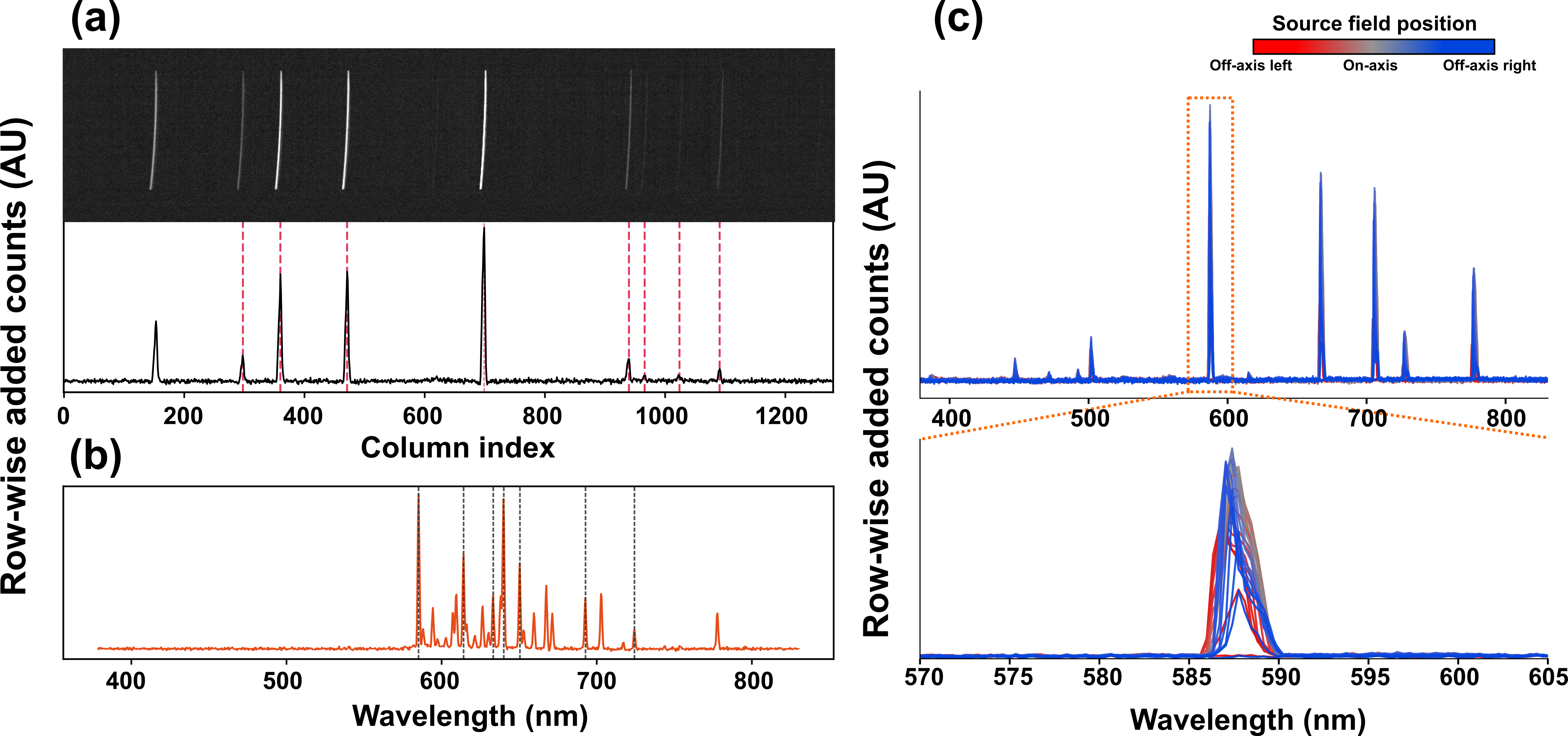}
\caption{Wavelength calibration. (a) Real photograph of helium's spectral image with its spectrum by column index used for calibration. (b) Neon's spectral image plotted using the calibration from the helium discharge lamp. The dashed lines represent the accepted values for seven isolated neon lines -- if our calibration is correct, then we should observe peaks at those locations, which we do. (c) Helium discharge lamp swept through the same FOV as Fig 2, maintaining the same wavelength-locking capabilities as the continuum light source.}
\end{figure}

Figure 3a shows the mapping from column index to wavelength. We use a helium discharge lamp on-axis to locate eight He I lines\textsuperscript{17} with sub-column precision across the visible wavelengths using a parabolic interpolation and apply a quadratic fit (\(\cos\theta\) dispersion variation expected across the detector), finding a mean reciprocal linear dispersion of 0.3533 nm/col (19.63 nm/mm at the sensor) with residuals 0.12 nm RMS (range over the calibrated band is 0.34-0.36 nm/col). Figure 3b tests the fit with a neon discharge lamp on-axis using the solution from the helium discharge lamp, agreeing to 0.13 nm RMS for seven isolated Ne I emission lines.\textsuperscript{17} No second-order features are observed in either the He or Ne lamp spectra as the angular Bragg selectivity of the VPH grating strongly suppresses orders other than the design order.\textsuperscript{18} Still, there is one line present near 777.2 nm that appears in both the He and Ne spectra which doesn't match either element's line list -- we attribute this to the O I triplet from trace oxygen in the lamps. We confirm this by the presence of the small, yet visible, peak at around column index 619 in the helium spectrum corresponding to wavelength 615.7 nm -- another O I triplet -- and by directly imaging the triplet structure at 777.2 nm (see Supplementary). Either way, such a contaminant does not affect the calibration nor the operation of the instrument and camera.

To obtain the final wander RMS error in terms of wavelength, the helium lamp is translated across the FOV where we track the motion of the 587.6nm spectral line. Figure 3c shows the final result: the spectral lines of helium remain locked up to 0.29 nm RMS, 1.03 nm peak-to-peak.
\Needspace{20\baselineskip}
\section{Spectral Resolution}
\begin{wrapfigure}{r}{0.5\textwidth}
\centering
\includegraphics[width=0.48\textwidth]{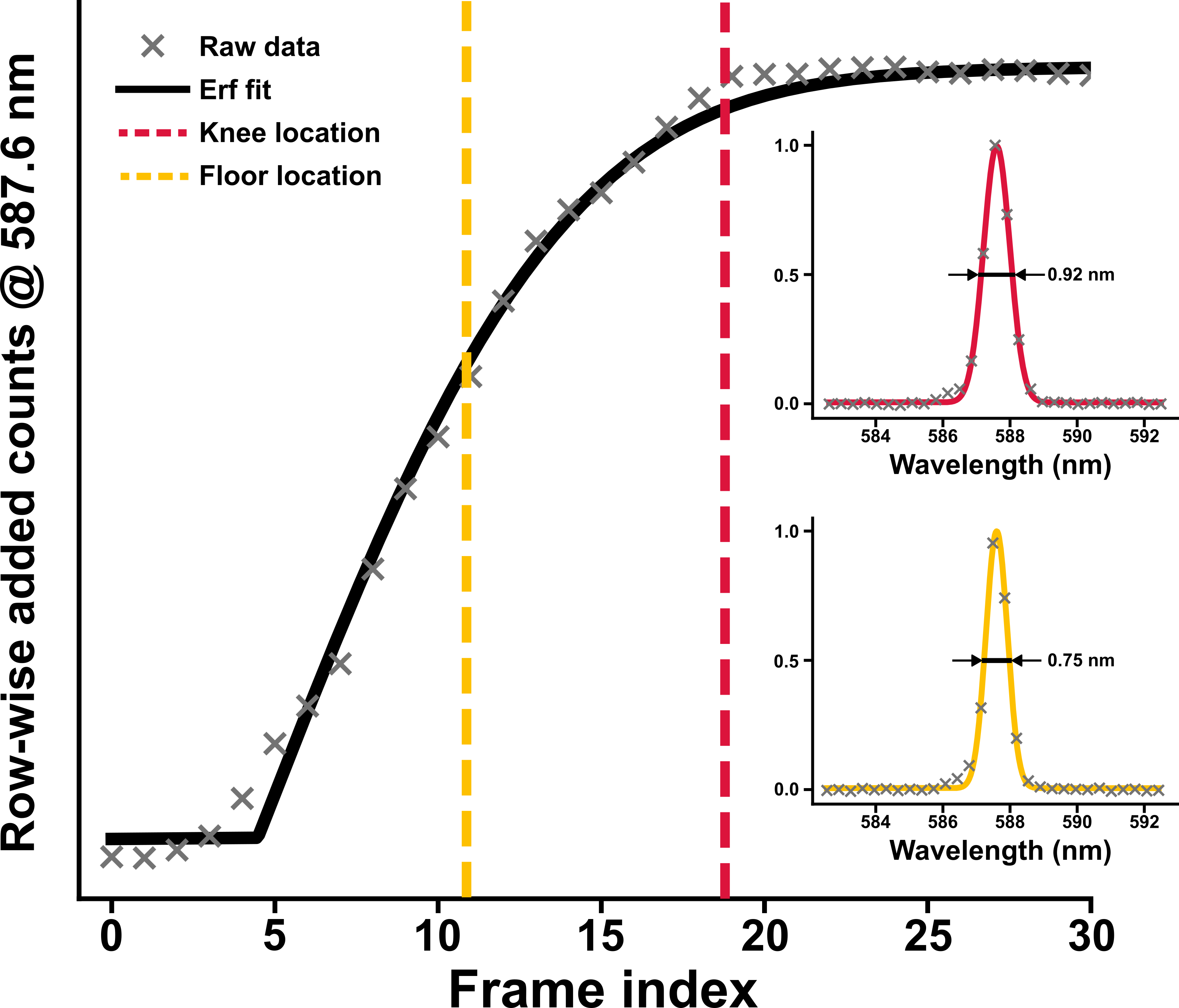}
\caption{Knee measurement. The slit aperture is opened continuously where each subsequent frame captures a new, wider slit width. This sweep is fit with the erf function. Insets: Gaussian fits for the He 587.6 nm line at the knee location (top) and at the floor location (bottom).}
\end{wrapfigure}
The spectral resolution is characterized by sweeping the width of the slit aperture and measuring the center-line counts of the He 587.6 nm line, shown in Figure 4. The slit aperture starts fully closed and is continuously swept until the center-line intensity no longer increases (opening the slit beyond this point adds light at new columns, not at the center column). At this point, the geometric slit image matches the optical blur, marking the slit width of maximum optical throughput without degrading the spectral resolution. We define the ``knee'' to be the operating point at 95\% of the saturated intensity. While this convolution model underlying this measurement is standard, we are not aware of a published procedure that sets the slit width through a continuous sweep and fitting the resulting intensity curve this way.

We assume a priori that the line-spread function of the image of the spectral lines is Gaussian. While the true profile is a convolution of the slit function, diffraction and aberration sources, and etc, the composite is still well-approximated by a Gaussian and the residuals of the following fits of this section are consistent with this assumption. Thus, the expected functional form of this continuous sweep is Gauss's error function \(\operatorname{erf}\) -- the integration of the Gaussian line-spread function over the slit width. With the slit opened at a constant rate, we have \(I(f) \propto \operatorname{erf}\left( \frac{\left( f - f_{0} \right)}{\tau} \right)\) where \(I\) is the intensity of the center column, \(f\) is the frame index, and \(f_{0}\) is the first frame at the onset of transmission. After fitting, we locate the knee and find the FWHM of the Gaussian 587.6 nm peak measured to be 0.92 nm (R $\approx$ 639 at 587.6 nm) while the median FWHM measured below the knee is 0.75 nm (R $\approx$ 783 at 587.6 nm). Therefore, the instrument resolves 0.92 nm at full throughput -- well below, for example, the potassium doublet separation important for the following demonstration.
\section{Demonstration}
\begin{figure}[!htb]
\centering
\includegraphics[width=\textwidth,height=0.88\textheight,keepaspectratio]{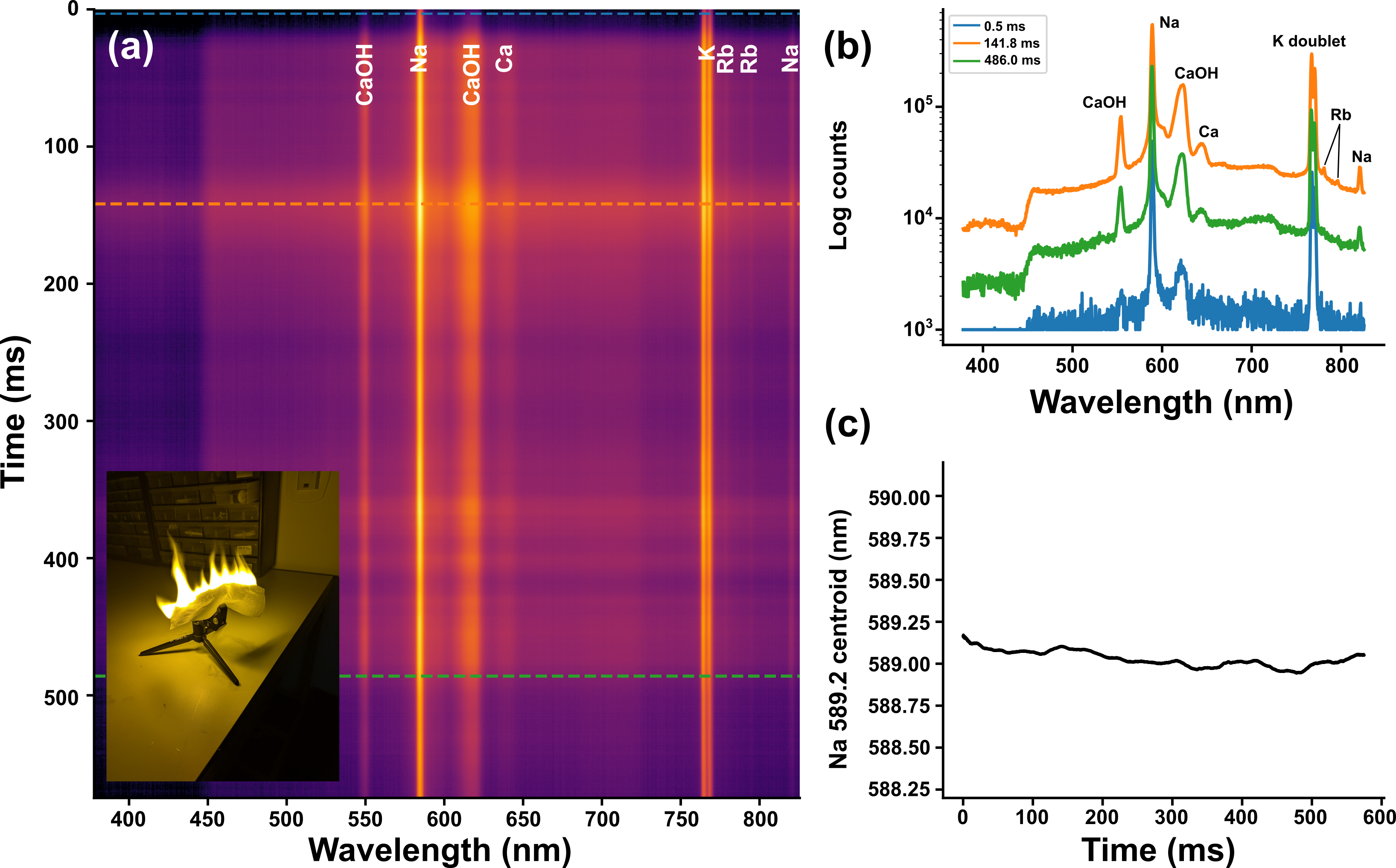}
\caption{Combustion spectroscopy. (a) Waterfall plot of intensities over time showing the chemical species CaOH, Na, Ca, K, and Rb. Inset: Photograph of the combusting nitrocellulose. (b) Log plot of the spectrum for three selected time slices. (c) Tracking the Na 589 nm centroid throughout the course of the combustion, confirming its stability at the sensor plane.}
\end{figure}

To demonstrate the properties of the high-speed camera and the instrument, nitrocellulose is placed at a 3m working distance from the 180mm fore-optic. When ignited, nitrocellulose produces uneven, spatially extended flame fronts that translate and move unpredictably. Figure 5 shows the time-resolved spectrum featuring the chemical species of Na I at approximately 589.0 nm (the doublet blends at this resolution) and 819.5 nm, CaOH green and orange bands at 554.0 nm and 622.0 nm respectively, Ca I at 644.0 nm, the resolved K I doublet at 766.5 nm and 769.9 nm, and even Rb I lines at 780.0 nm and 794.7 nm.\textsuperscript{17} Tracking the sodium line, the centroid remains consistent up to 0.06 nm RMS (0.28 nm peak-to-peak) over the course of the combustion event. It should be noted as well that overwhelmingly the residual variation here is dominated by a slow component that tracks the burn intensity consistent with self-absorption: as the flame's core emits light, the cooler, less-excited outer layers of sodium atoms absorb and re-emit that light in a way that can shift the apparent position of the centroid as the light intensity evolves. By contrast, frame-to-frame jitter (how much the centroid shifts frame-to-frame) is 0.003 nm RMS; the noise floor.
\section{Discussion}
The pointing invariance here is achieved by placing the slit aperture at the pupil conjugate along the dispersion axis, necessarily discarding spatial information along that axis -- all constituent wavelengths of the source, no matter where it may be laterally, map to the same column on the sensor. Thus, two emitters displaced laterally along that axis produce a single spectrum that's a superposition of both source objects with no means to distinguish between each. For a single evolving front, this is intentional and desirable, but compared to the conventional slit spectrograph where full spatial resolution is preserved at the slit plane, that same resolution is the price paid for a spectral image that does not move.

Now, perpendicular to the dispersion axis, the cylindrical doublet has no power so the relayed image plane \emph{does} remain where L2 forms it (at the front focal plane of the cylindrical doublet). Since the sensor plane is conjugate to the slit plane and the vertical image is focused \(2f_{c}\) = 100 mm ahead of the slit, there is a total blur at the slit plane of \(2Df_{c}/f_{L2} = D\) where \(D\) is the full diameter of the iris aperture. This is a key limitation of the current setup: the number of resolvable vertical emitters across the field is consequently set by the open diameter of the iris aperture, which directly trades with throughput. While in principle this axis can provide coarse localization of vertically-stacked emitters and this would be the obvious next characterization, it is not reported here.

The FOV reported here is set by a combination of the fore-optic and the displacements of the light in the back end. For the former, this is standard photographic principles: longer lenses provide higher magnification and thus smaller FOV -- ideal when the source is far away from the instrument. For the latter, L3 converts lateral field angle to lateral displacement, which the grating then adds a wavelength-dependent deviation on top of. These two displacements add, so extreme lateral off-axis field positions run out of camera lens pupil altogether and vignetting starts appearing at the blue end on one side of the field and the red end on the other side of the field. The 62 cm FOV at 1.5 m quoted here is therefore specific to the fore-optic (50mm f/1.2) and camera lens (85mm f/1.4) used for the sweep -- for a fore-optic agnostic quantity, we can express in terms of image height. An object that is 31 cm off-axis laterally imaged by a 50mm 1.5 m away lands \(h_{\max} = 10.7\ \mathrm{mm}\) off-axis at the internal image plane. With this quantity, we can check the usable FOV for other focal length photographic lenses: swap to the 180mm as we did for the experiment and the usable FOV is \(\frac{h_{\max}(L - f)}{f} = 10.7 \times \frac{2820}{180} = 16.8~\mathrm{cm}\) half-field so, 33.6 cm full. The nitrocellulose strip was 15.2 cm long, within the usable FOV.
\section{Conclusion}
We have demonstrated a spectrograph built from standard, stock optics around an unmodified high-speed camera with wavelength-locking capabilities decoupled from source position. The spectrograph features an internal telecentric stop and angle-space slit conjugate that ensures the spectral image does not wander as it illuminates the sensor. The instrument delivers R $\approx$ 639 at its operating slit width over the visible-NIR band with 0.12 nm RMS on-axis and 0.29 nm RMS across the whole field wavelength accuracy. As an example application, we ignited nitrocellulose and recorded the spatially extending flame front at 10 kHz, 99 \textmu{}s exposure, identifying five different species using the helium discharge lamp column-to-wavelength calibration. The natural next application is exploding wire and pulsed-plasma diagnostics where the source expands several mm during the event and necessitates frame rates on the order of 100 kHz or 1 MHz. The demonstrated invariance here translates well to such applications for Stark and Doppler analysis.\textsuperscript{19,20}

\section*{Conflict of Interest}
M.V. and K.D.G. are employees of Vision Research, Inc. (Wayne, NJ, USA), a business unit of AMETEK, Inc., which manufactures the Phantom high-speed cameras (VEO 1310 and C980) used in this work. This research was conducted as part of the authors\textquotesingle{} employment. The authors declare no other competing interests.
\clearpage
\section*{Supplementary Information}\subsection*{ABCD Matrix Formulation}
A ray can be uniquely defined by two elements: height \(x\) (i.e. its lateral displacement from the optical axis) and slope \(u\) (the direction it's heading). In the thin lens paraxial approximation, there are only two linear operations that exist to act on \((x,u)^{T}\ \): 1) free space propagation and 2) refractive response to a lens.

Free space propagation by distance \(d\)

\[\begin{pmatrix}
x' \\
u'
\end{pmatrix} = \begin{pmatrix}
1 & d \\
0 & 1
\end{pmatrix}\begin{pmatrix}
x \\
u
\end{pmatrix} = \begin{pmatrix}
x + d \cdot u \\
u
\end{pmatrix}\]

Refractive response to a lens with focal length \(f\)

\[\begin{pmatrix}
x' \\
u'
\end{pmatrix} = \begin{pmatrix}
1 & 0 \\
 - 1/f & 1
\end{pmatrix}\begin{pmatrix}
x \\
u
\end{pmatrix} = \begin{pmatrix}
x \\
u - x/f
\end{pmatrix}\]

The one exception is the VPH grating which, to first order, adds a wavelength-dependent slope offset:

\[u \rightarrow u + \delta(\lambda),\ \ \ \ \delta(\lambda) = \frac{m}{d_{g}\cos\beta_{0}}\left( \lambda - \lambda_{0} \right)\]

where \(d_{g}\) is the groove period, m is diffraction order, and \(\lambda_{0}\) is the Bragg wavelength.
\subsection*{Optical Train}
Step 0: fore-optic

The photographic fore-optic can be treated as an approximate thin-lens with focal length \(f_{N}\). Consider a source somewhere in object space at a distance \(L_{O}\) and transverse position \(X_{O}\). The image plane \(P_{I}\) lies at \(L_{i}\) as determined by the thin lens equation \(\frac{1}{L_{O}} + \ \frac{1}{L_{i}} = \frac{1}{f_{N}}\) and corresponding magnification \(m_{N} = - \frac{L_{i}}{L_{O}} = X_{i}/X_{O}\).

We parameterize \(u\) as \(u = (\rho - X_{O})/L_{O}\) where \(\rho\) is the height at which the ray crosses the lens (assuming \(\rho\) is within the acceptance cone of the lens).
  Propagate \(L_{O}\) to lens: \[\begin{pmatrix}
  X_{O} + L_{O}u \\
  u
  \end{pmatrix} = \begin{pmatrix}
  \rho \\
  \left( \rho - X_{O} \right)/L_{O}\ 
  \end{pmatrix}\]
  Lens action \(f_{N}\): \[\begin{pmatrix}
  \rho \\
  \frac{\left( \rho - X_{O} \right)}{L_{O}} - \frac{\rho}{f_{N}}
  \end{pmatrix} = \begin{pmatrix}
  \rho \\
   - \frac{X_{O}}{L_{O}} - \frac{\rho}{L_{i}}
  \end{pmatrix}\]
  Propagate \(L_{i}\) to \(P_{I}\): \[\begin{pmatrix}
  \rho - L_{i}\left( - \frac{X_{O}}{L_{O}} - \frac{\rho}{L_{i}} \right) \\
   - \frac{X_{O}}{L_{O}} - \frac{\rho}{L_{i}}
  \end{pmatrix} = \begin{pmatrix}
  m_{N}X_{O} \\
   - \frac{X_{O}}{L_{O}} - \frac{\rho}{L_{i}}
  \end{pmatrix} = \begin{pmatrix}
  X_{i} \\
   - \frac{X_{O}}{L_{O}} - \frac{\rho}{L_{i}}
  \end{pmatrix}\]
Remark: Note that \(\rho\) vanishes from the position of the imaged ray after the fore-optic -- \emph{all} rays from the same field point \(X_{O}\) in object space land at the same field point \(X_{i}\) in image space no matter the angle of incidence they enter the lens.

Now, let's consider the slope at \(P_{I}\). Writing in terms of \(X_{i}\):

\[- \frac{X_{O}}{L_{O}} - \frac{\rho}{L_{i}} = \frac{X_{i}}{L_{i}} - \frac{\rho}{L_{i}} = \frac{X_{i} - \rho}{L_{i}}\]

At \(\rho = 0\), the chief ray has a slope \(u_{c} = X_{i}/L_{i}\ \). With the acceptance angle determined by the numerical aperture (NA) of the lens, the general form of the ray after the fore-optic for the rest of the analysis can be expressed as follows:

\[\begin{pmatrix}
x_{0} \\
u_{0}
\end{pmatrix} = \begin{pmatrix}
X_{i} \\
u_{0}
\end{pmatrix}\] with \(u_{0} \in \left\lbrack u_{c} - NA,\ u_{c} + NA \right\rbrack\)

The subsequent analysis is to determine \emph{which} function of \(\left( x_{0},u_{0} \right)\) survives at each plane.

Step 1: First relay element L1

We place the first relay doublet L1 with \(f_{1} \equiv f\) such that \(x_{0}\) is at its front focal plane.
  Propagate \(f\): \[\begin{pmatrix}
  x_{0} + fu_{0} \\
  u_{0}
  \end{pmatrix}\]
  Lens action L1: \[\begin{pmatrix}
  x_{0} + fu_{0} \\
  u_{0} - \frac{x_{0} + fu_{0}}{f}
  \end{pmatrix} = \begin{pmatrix}
  x_{0} + fu_{0} \\
   - x_{0}/f
  \end{pmatrix}\]
Remark: \(u_{0}\) dependence is now gone from the slope, so every ray from any field point \(x_{0}\) now travels the same slope, i.e. they are collimated.

Step 2: Iris

The iris is placed at the back focal plane of L1.

2.1 Propagate \(f\) to iris: \[\begin{pmatrix}
x_{0} + fu_{0} - f \cdot x_{0}/f \\
 - x_{0}/f
\end{pmatrix} = \begin{pmatrix}
fu_{0} \\
 - x_{0}/f
\end{pmatrix}\]

Remark: Now the position has lost \(x_{0}\) dependence. This is the aperture stop of the system. \textbf{Note how this is the position where} \(\mathbf{u}_{\mathbf{0}}\) \textbf{and} \(\mathbf{x}_{\mathbf{0}}\) \textbf{swap roles, scaled by the focal length.} We will return to this after a brief transition through step 3.

Step 3: Second relay element L2

We place the second relay element L2 one focal length \(f_{2} \equiv f\) away from the iris. Note that the two relay doublets are identical in optical diameter and focal length.

3.1 Propagate \(f\) to L2: \[\begin{pmatrix}
fu_{0} - f \cdot x_{0}/f \\
 - x_{0}/f
\end{pmatrix} = \begin{pmatrix}
fu_{0} - x_{0} \\
 - x_{0}/f
\end{pmatrix}\]

3.2 Lens action L2: \[\begin{pmatrix}
fu_{0} - x_{0} \\
 - \frac{x_{0}}{f} - \frac{\left( fu_{0} - x_{0} \right)}{f}
\end{pmatrix} = \begin{pmatrix}
fu_{0} - x_{0} \\
 - u_{0}
\end{pmatrix}\]

3.3 Propagate \(f\) to back focal plane: \[\begin{pmatrix}
fu_{0} - x_{0} - fu_{0} \\
 - u_{0}
\end{pmatrix} = \begin{pmatrix}
 - x_{0} \\
 - u_{0}
\end{pmatrix}\]

Remark: The reader might be wondering what the purpose of all of this is -- it seems like the only thing the relay did was negate \(x_{0}\) and \(u_{0}\). Let's address this more specifically.

Interlude -- Telecentricity

At a general plane anywhere along the optical path, a ray arrives at a height \(\rho = Ax_{0} + Bu_{0}\) (with \(A\), \(B\) fixed by the elements in between) so an aperture at any arbitrary location transmits the slope window \(u_{0} \in - \left( \frac{A}{B} \right)x_{0} \pm R/|B|\) where \(R\) is the half-width of the aperture stop. \textbf{Note how the center of the stop wanders with the} \(\mathbf{x}_{\mathbf{0}}\) \textbf{position}, the fore-optic itself is the most offending example here -- see step 0. At the exit pupil, the ray height is \(\rho = X_{i} - L_{i}u_{0}\), so it admits \(u_{0} \in u_{c} \pm NA\) as discussed earlier -- this is an acceptance cone whose center \emph{tilts} based on the field point in object space.

Recall our goal: we want pointing invariance, a system where any light imaged by the fore-optic does not wander based on where the object is. Therefore, we want to place the aperture stop of the system such that the aperture selects for these privileged rays, specifically when \(A = 0\) such that the center of the acceptance cone does not depend on \(x_{0}\).

Consider what step 2.1 produced. A ray arrives at the iris plane as

\[\begin{pmatrix}
\ x\  \\
u
\end{pmatrix}_{iris} = \begin{pmatrix}
fu_{0} \\
 - x_{0}/f
\end{pmatrix} = \begin{pmatrix}
0 & f \\
 - 1/f & 0
\end{pmatrix}\begin{pmatrix}
\ x_{0} \\
u_{0}
\end{pmatrix}\]

This position exactly satisfies the condition -- an aperture stop here will select for these privileged rays where the center of the acceptance cone has \emph{zero} dependence on the field point in object space since \(A = 0\), as defined above. Explicitly, the set of \(u_{0}\) allowed through from the aperture stop at this location is \(u_{0} \in \pm R/f\). \textbf{Note the lack of any} \(\mathbf{x}_{\mathbf{0}}\) \textbf{dependence.}

Thus, without the iris aperture, the system would transmit collimated bundles centered on a slope proportional to the field point in object space and as such, the image at the slit plane would drift with field point. With the iris aperture, if it's stopped down sufficiently enough (specifically that it becomes the true aperture stop of the system), every field point imaged by the fore-optic will deliver the exact same identical bundle since it is not dependent on the field point -- it selects for these privileged rays. This is called \textbf{image-space telecentricity}.

Step 4: Cylindrical lens

We place the cylindrical lens with focal length \(f_{c}\) such that its front focal plane is at the back focal plane of L2. The slit is placed at the back focal plane of the cylindrical lens.

4.1 Propagate \(f_{c}\) to the lens: \[\begin{pmatrix}
 - x_{0} - f_{c}u_{0} \\
 - u_{0}
\end{pmatrix}\]

4.2a Lens action along power axis: \[\begin{pmatrix}
 - x_{0} - f_{c}u_{0} \\
 - u_{0} + \frac{x_{0} + f_{c}u_{0}}{f_{c}}
\end{pmatrix} = \begin{pmatrix}
 - x_{0} - f_{c}u_{0} \\
x_{0}/f_{c}
\end{pmatrix}\]

4.2b Lens action along no-power axis: Nothing, remains as \[\begin{pmatrix}
 - x_{0} - f_{c}u_{0} \\
 - u_{0}
\end{pmatrix}\]

4.3a Propagate \(f_{c}\) to the slit along power axis: \[\begin{pmatrix}
 - x_{0} - f_{c}u_{0} + f_{c} \cdot \frac{x_{0}}{f_{c}} \\
x_{0}/f_{c}
\end{pmatrix} = \begin{pmatrix}
 - f_{c}u_{0} \\
x_{0}/f_{c}
\end{pmatrix}\]

4.3b Propagate \(f_{c}\) to the slit along no-power axis: \[\begin{pmatrix}
 - x_{0} - {2f}_{c}u_{0} \\
 - u_{0}
\end{pmatrix}\]

Remark: Thus our final result for the power axis of the cylindrical lens

\[\left( \ \begin{matrix}
x \\
u
\end{matrix}\  \right)_{slit} = \begin{pmatrix}
 - f_{c}u_{0} \\
x_{0}/f_{c}
\end{pmatrix}\]

demonstrates that there is no \emph{explicit} dependence of \(x_{s}\) on the field point \(x_{0}\) -- but keep in mind that there still could \emph{implicit} dependence of \(x_{s}\) on the field point \(x_{0}\) based on the set of \(u_{0}\) that arrives at the slit. As mentioned during the interlude, if the aperture stop remained at the fore-optic's pupil, \(u_{0} \in \frac{x_{0}}{L_{i}} \pm NA\), so illumination would wander with pointing, being smuggled in by the set of \(u_{0}\) allowed through by the wrong aperture stop location. However, with the aperture stop placed at the correct location that only allows through \(u_{0} \in \pm R/f\), there is neither explicit nor implicit dependence on \(x_{0}\). The illumination is locked. If the fore-optic collects the light, it reaches the slit in exactly the same point.
\subsection*{Oxygen triplet}
\begin{figure}[!htb]
\centering
\includegraphics[width=0.85\textwidth,height=0.88\textheight,keepaspectratio]{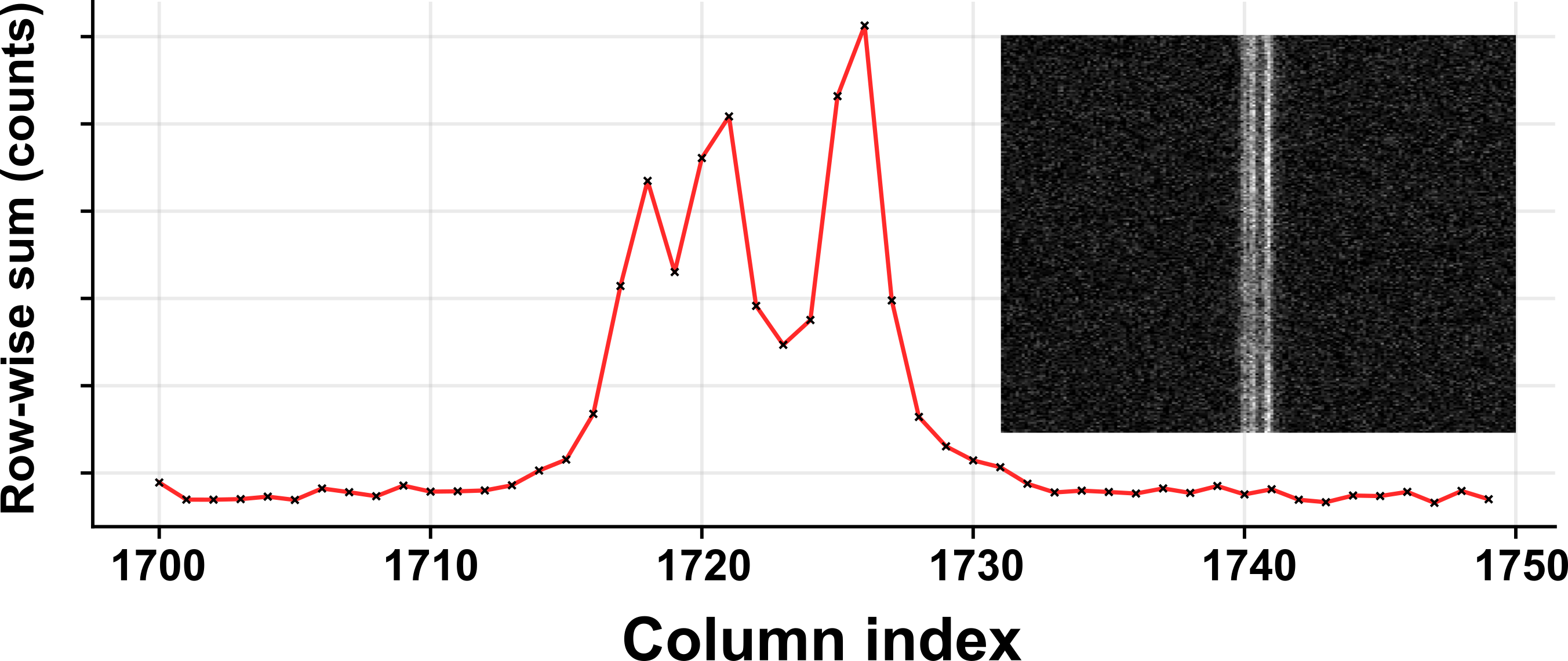}
\caption*{\textbf{Figure S1 --} Oxygen triplet confirmation. Each cross represents a column on the sensor. Inset: Photograph of the oxygen triplet showing its structure.}
\end{figure}

To confirm that there are absolutely no higher order diffraction modes polluting the first order spectrum, we replaced the 85mm camera lens with a 180mm camera lens, switched to a 5 \textmu{}m pixel pitch camera (Phantom C980), closed down both the slit aperture and iris aperture as far as they could go without vignetting the light, and zoomed into the unknown spectral line around 777 nm. Figure S1 shows the triplet structure with peak-to-peak spacing of 3 columns and 5 columns (each column value shown with a cross). The ratio between the peak-to-peak spacing in terms of wavelength is 0.223 nm / 0.122 nm = 1.8, here it's 5 col / 3 col = 1.7. This triplet structure coupled with the extra peak at 615 nm where another oxygen triplet is expected as well as the same feature appearing in the neon spectrum confirms that this is oxygen, not second-order blue features appearing at the sensor.

\end{document}